\documentclass[final,5p,times,twocolumn]{elsarticle}

\usepackage[T1]{fontenc}
\usepackage[utf8]{inputenc}
\usepackage{amsmath,amssymb,bm}
\usepackage{graphicx}
\usepackage{booktabs}
\usepackage{siunitx}
\usepackage{microtype}
\usepackage{hyperref}

\hypersetup{colorlinks=true,linkcolor=blue,citecolor=blue,urlcolor=blue}
\graphicspath{{figures/}}
\journal{Physics Letters A}
\newcommand{\ket}[1]{|#1\rangle}
\newcommand{\bra}[1]{\langle #1|}

\begin{document}

\begin{frontmatter}

\title{Direct imaging of Su-Schrieffer-Heeger edge modes in a disordered suspended mechanical oscillator chain}

\author[mcgill]{Defeng Lu}
\author[mcgill]{Rohan Modi}
\author[mcgill]{Nicolas Delnour}
\author[mcgill]{Michael Hilke}
\address[mcgill]{Department of Physics, McGill University, Montréal, Quebec H3A 2T8, Canada}

\begin{abstract}
We report a reconfigurable macroscopic mechanical realization of the Su-Schrieffer-Heeger (SSH) model: a suspended chain of alternating masses coupled by identical compression springs. The relevant SSH degree of freedom is the relative longitudinal displacement between neighboring masses. In this bond-coordinate basis, applying a constant shift and rescaling of the mechanical dynamical matrix maps a chain of $N$ blocks onto an $(N-1)$-site SSH chain. Using video tracking, we reconstruct bond-resolved spectra and mode profiles and compare them with predictions for an ideal finite SSH chain. The topological termination supports mid-gap modes localized near the boundaries, with the alternating-sublattice structure expected for an effective even-length SSH chain. Mass variations split the edge-mode frequencies and produce asymmetric mode profiles relative to the non-disordered predictions. This directly imageable platform enables studies of topological normal modes and disorder effects in a classical mechanical system.

\end{abstract}


\end{frontmatter}

\section{Introduction}
The Su-Schrieffer-Heeger (SSH) chain is the minimal lattice model exhibiting a one-dimensional topological band structure. It was developed to describe solitons in polyacetylene, where alternating bond strengths define two inequivalent dimerizations \cite{Su1979,Heeger1988}. In the tight-binding representation, the two alternating coupling patterns can have identical bulk energy spectra and band gaps but different winding numbers. For a finite open chain, the topological dimerization gives rise to boundary-localized zero modes, providing a simple manifestation of bulk-boundary correspondence in a chiral-symmetric system \cite{Ryu2002,Asboth2016}. Moreover, topological invariants can remain robust against disorder \cite{Jezequel2022}, which may also preserve topological transport signatures \cite{Ostahie2021}. The simplicity of the model also allows it to be extended to investigate topological-insulator behavior in higher dimensions \cite{Li2022,Zhou2018}. These properties make the one-dimensional SSH model a promising platform for experimental realizations and have established it as a standard reference for real-space probes of topology in diverse contexts \cite{Leder2016,Zaimi2021,Fan2022}.

Mechanical and acoustic systems provide a particularly transparent route to topological band physics because their classical mode profiles and local phases can be directly measured through displacement fields \cite{Huber2016,Sustrunk2016, Yang2015}. This direct access has motivated extensive work on topological phenomena in mechanical systems \cite{Kane2014,Xiao2015, Roman2015}. Compared with microscopic quantum systems, mechanical platforms offer direct real-space access to mode amplitudes and phases, facilitating controlled studies of perturbations such as geometric disorder, damping, and coupling to unwanted degrees of freedom \cite{Snee2019,Chaunsali2021}.

Although the one-dimensional SSH model has been realized in many different media \cite{Neder2024,Thatcher2022,Coutant2021}, a natural macroscopic mechanical analogue is a one-dimensional mass-spring chain \cite{Sircar2025}. Experiments on mechanical systems have demonstrated tunable topological transitions and boundary-localized modes, both valuable features for investigating topological models \cite{Chaunsali2017,kuo2023sshcoupledspringsystems,ozdemir2023rigorous}. Alternating-mass chains coupled by identical springs have also been shown to exhibit SSH topology when described in strain or relative-displacement coordinates \cite{Allein2023}. Analytical and numerical studies further indicate that these systems can remain robust in the presence of disorder \cite{Shi2021,Carhart_2024}.

To investigate how mass disorder affects a mechanical SSH analogue, we developed a suspended chain of alternating masses coupled by identical springs. In the bond-coordinate representation, the alternating masses generate the effective couplings required by the SSH model. Thus, a system of $N$ oscillators maps onto an $(N-1)$-site SSH chain. Mass variations introduce disorder in these couplings.  Rearranging the masses allows us to realize both SSH dimerizations. We use video tracking and spectral reconstruction to identify edge-localized mid-gap modes in the topological configuration. We first compare the measured modes and mode profiles with the disorder-free theoretical finite-chain SSH eigenvectors, then characterize deviations from this ideal case by computing an effective two-level system, which correctly captures the effect of disorder on the edge states.

\section{Exact mapping of the mass chain to the SSH model}

To map a mechanical analogue onto the SSH model, we consider a one-dimensional chain of $N$ blocks connected by identical springs of stiffness $k$.  The longitudinal displacement of block $i$ is $x_i$, and the masses $m_i$ can be different.   For an interior block, the motion is described by 
\begin{equation}
  m_i\ddot{x}_i = k(x_{i-1}-2x_i+x_{i+1}),
\end{equation}
with the corresponding one-sided equation for the free ends. Combining the equations of motion of all blocks, we obtain $\ddot{\bm{x}}=D\bm{x}$, where $D$ is the displacement-space dynamical matrix that contains different mass factors on the diagonal and adjacent off-diagonal entries.  The absolute-displacement matrix is not itself equivalent to the SSH Hamiltonian with alternating masses. However, when we look at the relative displacement 
\begin{equation}
  q_i=x_i-x_{i+1},\qquad i=1,\ldots,N-1 ,
  \label{eq:bondcoords}
\end{equation}
we can write the corresponding dynamical equation as $  H_q\bm{q}=\omega^2\bm{q}$ where $\omega^2$ are the eigenfrequencies associated with the wavefunctions $q$, with

\begingroup
\setlength{\arraycolsep}{1.6pt}
\begin{equation}
  H_q=k\begin{pmatrix}
  (m_1^{-1}+m_2^{-1})&-m_2^{-1}&0&\cdots\\
 -m_2^{-1}&  (m_2^{-1}+m_3^{-1})&-m_3^{-1}&\cdots\\
  0&-m_3^{-1}& (m_3^{-1}+m_4^{-1})&\cdots\\
  \vdots&\vdots&\ddots&\ddots\\
  0&\cdots&-m_{N-1}^{-1}& (m_{N-1}^{-1}+m_N^{-1}).
  \end{pmatrix}
  \label{eq:sshmatrix}
\end{equation}
\endgroup

For alternating masses $m_1$ and $m_2$ and the following transformation, we recover the conventional dimensionless SSH model \cite{Alseiari2023,alseiari2023thermalrectificationtopologicaledge}:
\begin{equation}  
\begin{aligned}
H_{SSH}&=\left(\sqrt{\frac{m_1}{m_2}}+\sqrt{\frac{m_2}{m_1}}\right)\mathbb{I}-\sqrt{m_1m_2}\,\frac{H_q}{k} \\
&=
\sum_{n=1}^{N-2} \left(t_n b_{n+1}^\dagger b_n+t_n b_n^\dagger b_{n+1}\right),
 \label{eq:shifted}
\end{aligned}
\end{equation}
where $t_{2n-1}=\sqrt{m_1/m_2}$ and $t_{2n}=\sqrt{m_2/m_1}$. For a chain of $N$ masses, the equivalent SSH model contains $N-1$ sites described by the $N-1$ annihilation and creation operators  $b_{n}$ and $b_n^\dagger$, respectively. An interesting consequence of this mapping, relevant to our experimental implementation, is that even in the presence of randomness in the masses, the model maps exactly onto the disordered SSH model \cite{hilke2026disordered}, 
\begin{equation}  
H_{D-SSH}=
\sum_{n=1}^{N-1}v_nb_{n}^\dagger b_n+\sum_{n=1}^{N-2} \left(t_n b_{n+1}^\dagger b_n+t_n b_n^\dagger b_{n+1}\right),
 \label{eq:disordeed}
\end{equation}
where $v_n= \sqrt{\bar{m}_1\bar{m}_2 }\left[\bar{m}_1^{-1}+\bar{m}_{2}^{-1} -(m_n^{-1}+m_{n+1}^{-1})\right]$ is the onsite disorder of average zero and $t_n=\sqrt{\bar{m}_1\bar{m}_2}\,m_{n+1}^{-1}$ the hopping energy. $\bar{m}_1$ ($\bar{m}_2$) is the average of the odd (even) mass. Here we neglect the difference between the inverse of the average mass and the average of the inverse mass. In this disordered model, there are correlations in the disorder between the onsite and hopping disorder as well as between neighboring onsite potentials, which we neglect. 
The experimentally measured "wavefunction" is the vector of bond oscillations $\bm{q}$ rather than the vector of block displacements $\bm{x}$. The measured resonance frequencies of the mechanical analogue system correspond to the "energy levels" of the SSH model.

The system can be configured into a topologically trivial or non-trivial configuration based on how the hopping is terminated. Since $t_n\sim m_{n+1}^{-1}$, a light intermediate mass produces a strong hopping and a heavy intermediate mass produces a weak hopping. The non-trivial (topological) termination begins and ends with weak hoppings. This supports mid-gap modes localized near the chain ends. Conversely, the trivial termination begins and ends with strong hoppings, and does not support mid-gap modes.  For an effective odd-length SSH chain, chiral symmetry enforces a single mid-gap state in the ideal model, whose amplitude occupies only one sublattice. For an effective even-length chain, the two edge states hybridize into symmetric and antisymmetric combinations with an exponentially small splitting. Here we consider only consider the even effective SSH chain.

\section{Experimental implementation and motion analysis} \label{Methods}

\begin{figure}[t]
\centering
\includegraphics[width=0.95\linewidth]{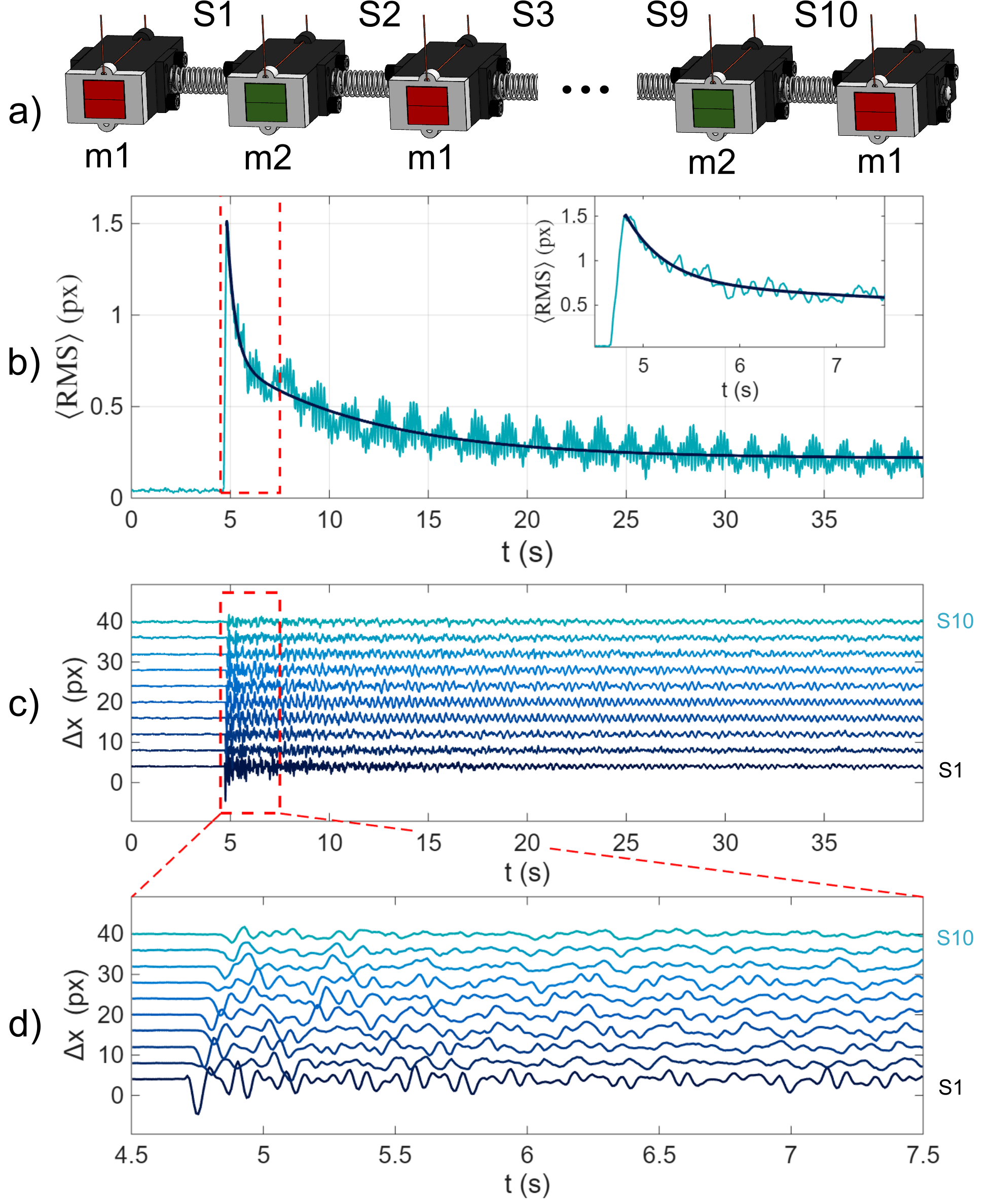}
\caption{Design and Signal Processing: (a) Design graphic of the spring-mass coupled pendulum system with springs labeled from left to right and the masses alternating. (b) Average RMS displacement of window 0.125s across all 10 springs over a 40-second acquisition period, with an inset showing the time close to the impact. The parameters extracted from the double exponential decay curve in Eq, \eqref{rms} are $(x_0=0.217 \pm 0.002) \text{px}$, $A_1=(0.535 \pm 0.007)\text{px}$, $\lambda_1=(0.139 \pm 0.003)\text{Hz}$, $A_2=(0.76 \pm 0.02)\text{px}$, $\lambda_2=(2.5 \pm 0.1)\text{Hz}$. (c) Time-series displacement of the 10 springs, with (d) an expanded view of the selected time period around the impact.}

\label{fig:timeseries}
\end{figure}

As shown in Fig.~\ref{fig:timeseries}(a), the system consists of $N=11$ oscillators connected by compression springs and suspended from the ceiling by wires to minimize friction during longitudinal motion. Each oscillator consists of a hollow, rectangular 3D-printed holder containing a colored mass cartridge whose color identifies the oscillator as light or heavy. Spherical brass inserts can be added to each cartridge to adjust the mass and thereby tune the resonance-frequency range. Each cartridge is held in place by friction, allowing the mass arrangement to be readily reconfigured between the trivial and topological configurations of the mechanical SSH analogue. The suspension wires have a length of $(1.500\pm0.003)\text{m}$, the springs have an effective stiffness of $k=(191\pm3)\frac{\text{N}}{\text{m}}$ 
and the light and heavy masses of $(23.6\pm 3.5)\text{g}$ and $(59.3\pm 5.9)\text{g}$, respectively. Because the per-site mass variance is more important than that in the spring stiffness, we assume that mass variations are the dominant source of disorder represented in Eq.~\eqref{eq:disordeed}.

The system is excited using two methods. First, a light tap is applied to the system to approximate a delta-function impulse and excite the full resonance spectrum. Second, an oscillator at one free end is driven continuously and weakly at a selected resonance frequency to isolate the corresponding mode. The pendular suspension inevitably introduces additional pendular, bending, and yaw degrees of freedom that lie outside the one-dimensional model and can project onto the measured longitudinal coordinate. To minimize these contributions and isolate the longitudinal displacement as effectively as possible, the system is recorded from below. Videos are acquired at 120 frames per second, allowing frequencies up to the Nyquist limit of \SI{60}{\hertz} to be resolved. The system parameters were chosen so that all resonance frequencies lie below this limit. The videos have a longitudinal resolution of 1980 pixels, and the chain length was selected to balance resonance-frequency resolution with an effective longitudinal resolution of $89\pm4$ pixels per oscillator.

After video tracking, the block displacements are converted into bond displacements. Fig.~\ref{fig:timeseries} (b)-(d) shows the average root-mean-square (RMS) displacement, the time series for each spring, and an enlarged view of the response near the impact. The excitation response is clearly visible in these panels. For times after the impact time of $(t_0=4.81 \pm 0.01)\, s$, the RMS displacement is fitted by a double-exponential decay

\begin{equation}
    \text{RMS}(t)=x_0+\Theta(t-t_0)\sum_{n=1}^2A_ne^{-\lambda_n(t-t_0)}.
    \label{rms}
\end{equation}

The exponential decays are associated with the longitudinal motion and 2 dominant damping mechanisms, while the persistent periodic motion is attributed mainly to the pendular motion of the system. The different decay times help distinguish the longitudinal response from other degrees of freedom and supports focusing the subsequent analysis on the longitudinal component. The steep initial decay indicates substantial dissipation of the longitudinal motion and limits the duration over which its ringdown can be reliably measured. A \SI{40}{\second} measurement window therefore balances the need to capture the relevant transient response with the need for sufficient frequency resolution. The complex spring-motion time traces indicate the simultaneous excitation of multiple resonance modes, with amplitudes that vary from spring to spring. These spatial variations are reflected in the corresponding mode profiles.

\section{Spectra and wavefunctions}

\subsection{Delta impulse excitation}

\begin{figure*}[t]
\centering
\includegraphics[width=0.95\linewidth]{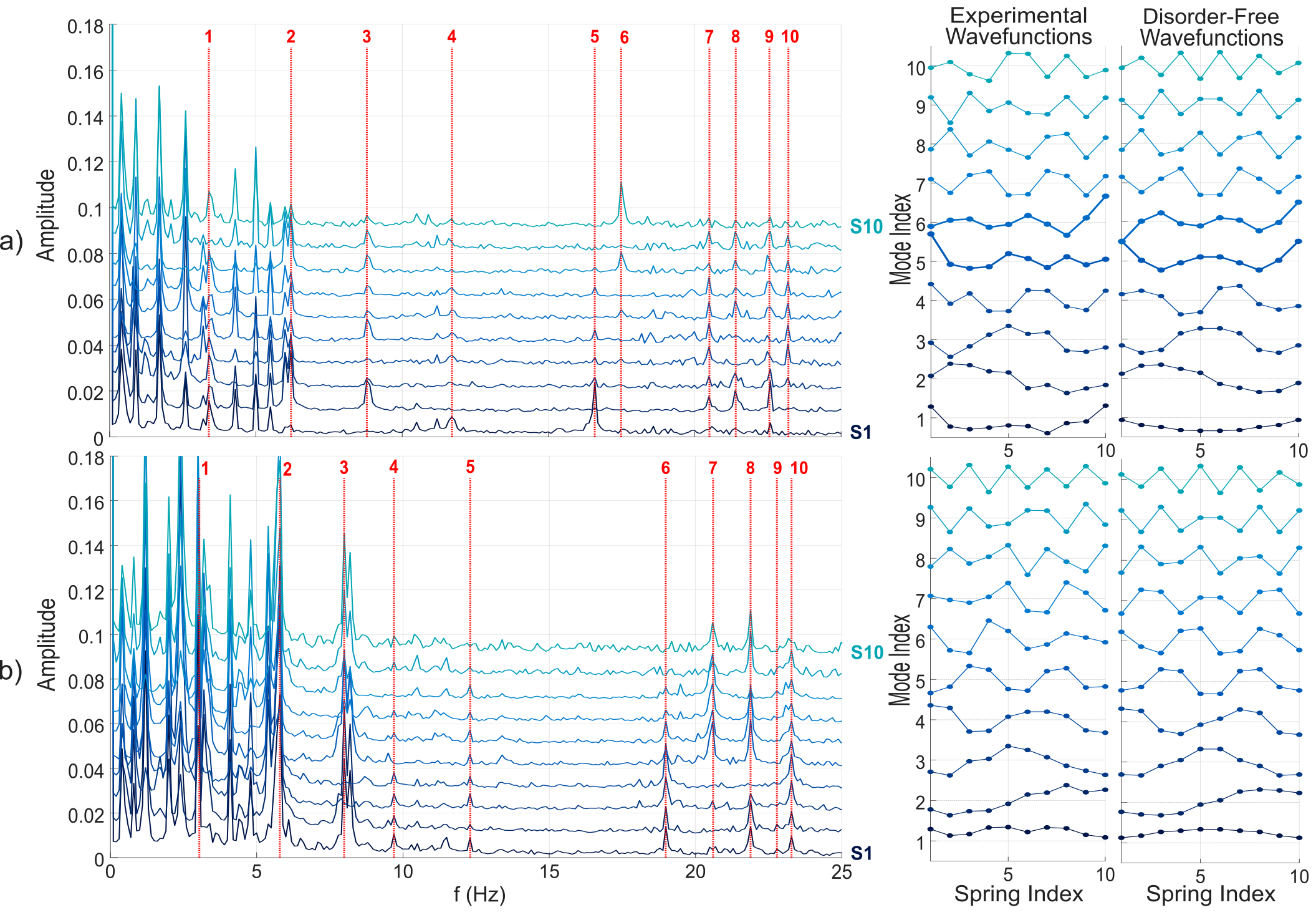}
\caption{Frequency spectra for all 10 springs with the located peak frequencies labeled by numbered red lines and a comparison between the experimentally extracted peak frequency wavefunctions with the disorder-free theoretical wavefunctions for an ideal SSH system for (a) the topological configuration and (b) the trivial configuration. To maximize visibility, a higher sliding FFT averaging window was used for the lower frequencies to minimize noise and a lower one for the higher frequencies where the spectra were weaker to highlight the signal. These separation points were 15Hz for the topological configuration and 18Hz for the trivial configuration.}
\label{fig:spectrum}
\end{figure*}

\begin{figure}[t]
\centering
\includegraphics[width=0.80\linewidth]{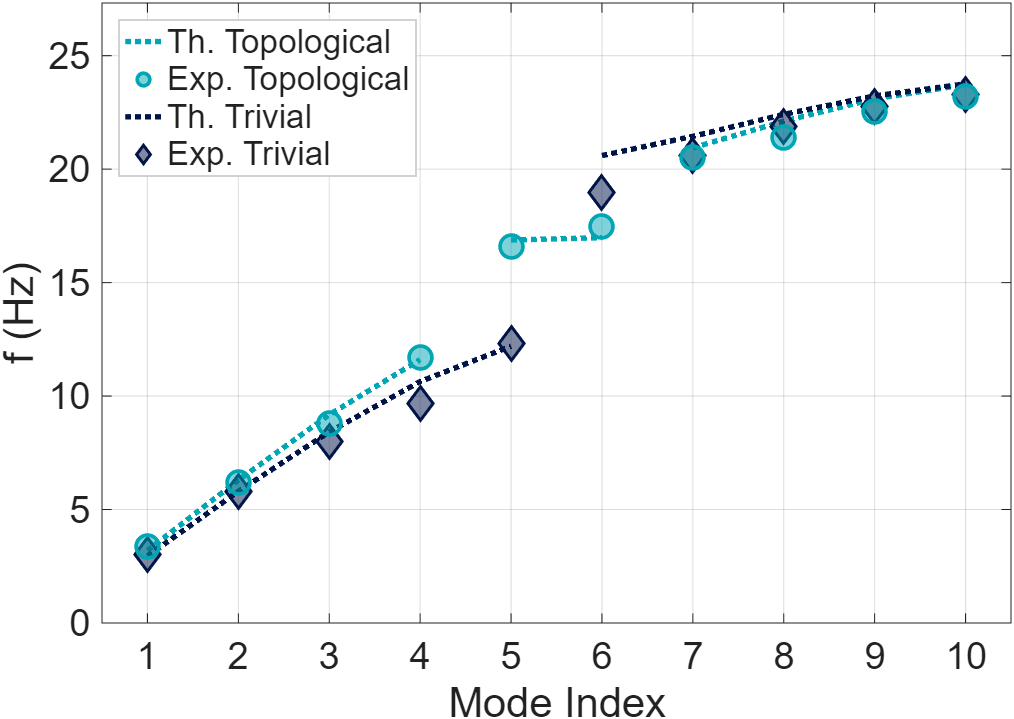}
\caption{Experimental and theoretical discrete resonance frequency spectrum plot for both topological and trivial configurations. The theoretical spectrum considers the system as an ideal, disorder-free analogue of the SSH model using the dimensions reported in Section 3.}
\label{fig:scatter}
\end{figure}

Using the first excitation method described in Section~\ref{Methods}, we extracted the resonance-frequency spectra of the mechanical SSH analogue in both the topological and trivial configurations, as shown in the two leftmost panels of Fig.~\ref{fig:spectrum} (a) and (b) respectively. At lower frequencies, neighboring peaks overlap because of pendular motion and the projection of other modes onto the longitudinal coordinate. At higher frequencies, the weaker spectral response results in a lower signal-to-noise ratio. We therefore used different averaging-window lengths in the fast Fourier transform (FFT) analysis to improve the visibility of candidate resonance peaks. These peaks were initially identified using the disorder-free theoretical resonance frequencies in Fig.~\ref{fig:scatter} as references. The robustness of each nearby spectral peak was then evaluated through a window-length sensitivity analysis of the short-time Fourier transform. Peaks that persisted over a range of window lengths were classified as robust stationary resonance modes, whereas transient and noise-induced features were attenuated by increased temporal averaging. The spatial profiles of the corresponding wavefunctions provide an additional validation of the selected resonance frequencies.

The experimental wavefunctions shown in the two middle panels of Fig.~\ref{fig:spectrum} were obtained from the real part of the FFT coefficient at each spring index, retaining both its amplitude and sign, and then normalizing the resulting vector to unit length. These profiles are compared with the theoretical wavefunctions of a disorder-free mechanical SSH analogue in the two rightmost panels. For both configurations, the agreement is stronger for the lower-frequency acoustic modes; nevertheless, the principal features of most higher-frequency optical modes remain visible in the experimental data. The resonance-frequency comparison in Fig.~\ref{fig:scatter} also shows reasonable agreement between the measured frequencies and the theoretical predictions indicated by the dotted lines.

The principal features of the SSH model, including the two edge states in the topological configuration and the clear band gap in the trivial configuration, are evident in both the wavefunctions and the frequency spectra shown in Figs.~\ref{fig:spectrum} and~\ref{fig:scatter}. However, notable discrepancies remain: the experimental band gap is slightly narrower, the two edge-mode frequencies are more widely separated, and each edge mode is localized predominantly on one side of the system. These deviations are consistent with disorder in the mechanical analogue, as discussed in section 4.3.

\subsection{Harmonically driven excitation}
\begin{figure}[t]
\centering
\includegraphics[width=0.95\linewidth]{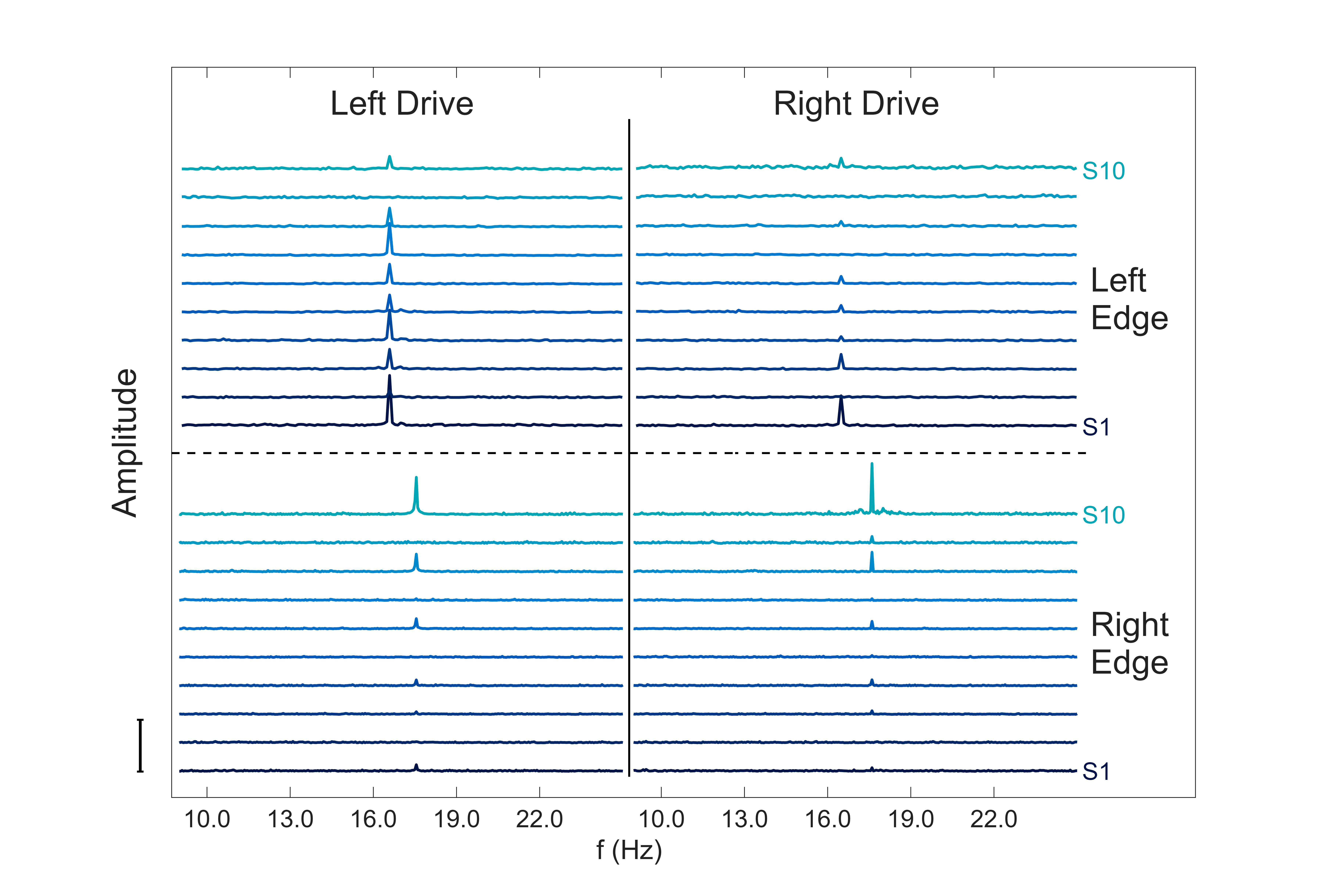}
\caption{Comparison of the edge mode amplitudes for all 10 springs resulting from the system being driven at the respective resonance frequency from the left and from the right. The scale bar indicates the normalized maximum amplitude in each column.}
\label{fig:peaks}
\end{figure}

To further investigate the effects of disorder on the nominally degenerate edge states, a defining feature of the SSH model, we use the second excitation method described in Section~\ref{Methods}. The system is driven acoustically via weak coupling from each end at the measured resonance frequency of each edge state, allowing the two modes to be isolated and analyzed separately. These resonance frequencies are respectively $f_5=(16.58 \pm 0.05)\, \text{Hz}$ and $f_6=(17.48\pm 0.05) \, \text{Hz}$. The main result shown in Fig.~\ref{fig:peaks} is spatially selective edge-mode excitation. Driving the system from the left boundary excites the left-localized mode more efficiently than driving it from the right boundary. Conversely, the right-localized mode is excited more efficiently from the right boundary. This edge-dependent selectivity contrasts with the symmetric response expected for an ideal mechanical SSH analogue. The alternating amplitude of the edge mode is also a key signature.

\subsection{Energies and eigenfunctions of the topological edge states in the presence of disorder}

The two main differences between the experimental energies and the theoretical clean SSH energies are (a) the energy difference and (b) asymmetry in the edge state wavefunctions. This can be understood quite naturally in the context of the disordered SSH model, as defined in Eq.\eqref{eq:disordeed}. In the absence of disorder, the topological edge states can be obtained from the hybridization of the left and right edge basis states given by
\begin{equation} 
\begin{aligned}
\ket{L}=
C\sum_{n=1}^{N_S/2}(-r)^{n-1}\ket{2n-1} \text{ and }
\ket{R}=
C\sum_{n=1}^{N_S/2}(-r)^{N_S/2-n}\ket{2n},
\end{aligned}
\end{equation}
where $N_S=N-1$ is the number of sites ($N_S$ is even), $r= t_1/t_2$ with $r < 1 $ for the topological phase, and the normalization 
$C^2=(1-r^2)/(1-r^{N_S})$. The corresponding disorder-free Hamiltonian in the basis $\ket{L}$ and $\ket{R}$ is 
\begin{equation}
H_{\rm edge}^{(0)}=
\begin{pmatrix}
0 & \tau\\
\tau & 0
\end{pmatrix},
\end{equation}
with
\begin{equation}
\tau=\bra{L}H_{\text{SSH}}\ket{R}=
(-1)^{N_S/2-1}t_2
\frac{(1-r^2)r^{N_S}}{1-r^{N_S}}.
\end{equation}
$H_{\text{SSH}}$ is the disorder-free SSH Hamiltonian. The energy splitting is $2|\tau|$, which is the usual finite-size hybridization of the two SSH edge modes \cite{Su1979, Asboth2016,Chen2019}.

In the presence of independent onsite disorder $v_n$, with $\langle v_n \rangle=0$ and $\langle v_n^2 \rangle=V^2/12$ for a uniform distribution of width $V$, we project the disordered SSH Hamiltonian onto the states  
$\ket{L}$ and $\ket{R}$, which yields
\begin{equation}
H_{\rm eff}=
\begin{pmatrix}
v_L & \tau\\
\tau & v_R
\end{pmatrix},
\end{equation}
with effective onsite energies
\begin{equation}
\begin{aligned}
v_L=
C^2\sum_{n=1}^{N_S/2}r^{2(n-1)}v_{2n-1}\text{ and }
v_R=
C^2\sum_{n=1}^{N_S/2}r^{2(N_S/2-n)}v_{2n}.
\end{aligned}
\end{equation}
Their variances are given by 
\begin{equation}
    \langle v_L^2 \rangle=\langle v_R^2 \rangle=\langle v_n^2 \rangle \,\text{IPR},
\end{equation}
where the edge-state Inverse Participation Ratio (IPR) \cite{PhysRevLett.84.3690} yields 
\begin{equation}
    \text{IPR}\equiv\sum_{n=1}^{N_S}|\psi_{L,R}(n)|^4=
    \frac{1-r^2}{1+r^2}
    \frac{1+r^{\frac{N_S}{2}}}{1-r^{N_S}}.
\end{equation}
$\psi_{L,R}$ are the amplitudes of edge states $\ket{L}$ or $\ket{R}$. 
Since $v_L$ and $v_R$ are independent,
\begin{equation}
    \langle (v_L-v_R)^2\rangle=\sigma^2,
\end{equation}
where, $\sigma^2=2\langle v_n^2\rangle\,\text{IPR}$. Therefore, the variance of the energy difference is 
\begin{equation}
    \langle (E_L-E_R)^2\rangle
    =
    4\tau^2+\sigma^2.
    \label{eq.IPR}
\end{equation}

\begin{figure}[h]
\centering
\includegraphics[width=0.75\linewidth]{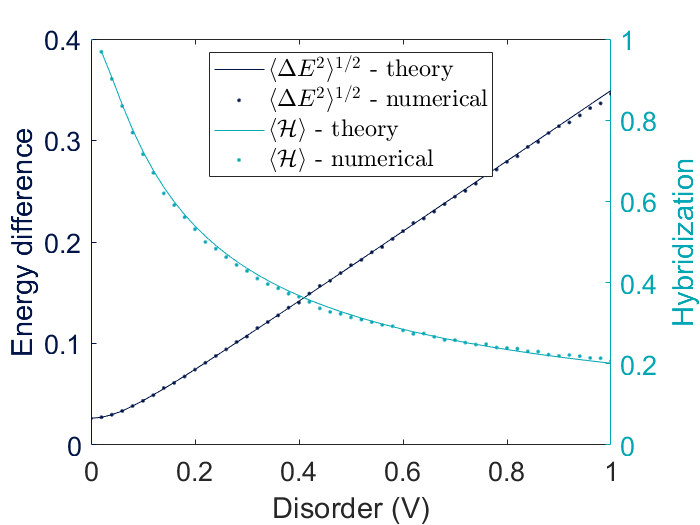}
\caption{Left axis: Energy difference $\langle [E_L-E_R]^2 \rangle^{1/2}$ as a function of the disorder strength amplitude $V$. The theory is obtained from Eq. \eqref{eq.IPR}. Right axis: Left and right edge hybridization as a function of $V$, where the theoretical curve is from Eq. \eqref{eq.hybrid}. The dotted points are the numerical values obtained by averaging over 10000 configurations with $N_S=10$, $t_1=0.63$, and $t_2=1/t_1$. Numerically, the average hybridization, $\langle\mathcal H\rangle$, of the two edge states ($\psi_\pm$) is evaluated using the ratio at the end sites $\left|\frac{\psi_\pm(1)}{\psi_\pm(N)}\right|$, yielding 
$
 \langle \mathcal H\rangle_{\text{num}} =
\left\langle
\operatorname{sech}\left[
\frac{1}{2}\left(
\left|\ln\left|\frac{\psi_-(1)}{\psi_-(N)}\right|\right|
+
\left|\ln\left|\frac{\psi_+(1)}{\psi_+(N)}\right|\right|
\right)
\right]
\right\rangle $.} 
\label{fig:disorder}
\end{figure}

The first term is the disorder-free edge-edge hybridization; the second is the edge-level detuning due to the disorder and is also the scale of disorder broadening of the
near-zero DOS. We verified numerically that the average energy separation is indeed consistent with expression \eqref{eq.IPR} as shown in Fig. \ref{fig:disorder}. The hybridization of the left and right edge states is given by $\mathcal H=\tau/\sqrt{\tau^2+(v_L-v_R)^2/4}$. For $v_L=v_R$ (no disorder) $\mathcal H=1$, i.e., the two edge states are fully hybridized with wavefunctions $\psi\sim|L\rangle\pm|R\rangle$. We can now estimate the expected energy splitting and hybridization of the edge states using our values of disorder. For our measured values of $k=(191\pm3)\frac{\text{N}}{\text{m}}$, $m_1=(23.6\pm\, 3.5)\text{g}$ and $m_2=(59.3\pm\, 5.9)\text{g}$, neglecting the error in $k$, and neglecting off-diagonal disorder, these values translate into an onsite disordered SSH model, with parameters $t_1=0.63$, $t_2=1/t_1$, and $\langle v_n^2 \rangle=\bar{m}_1\bar{m}_2(dm_1^2/\bar{m}_1^4+dm_2^2/\bar{m}_2^4)\simeq 0.059$, where $dm_\sigma$ is the standard deviation of $m_\sigma$. Using these values, the average splitting of the edge state energies is $\Delta E = \langle [E_L-E_R]^2 \rangle^{1/2}=\sqrt{4\tau^2+\sigma^2} \simeq 0.29$ and the hybridization averaged over disorder is
\begin{equation}\langle\mathcal H\rangle_\text{th}
    =\frac{2\tau}{\sqrt{2\pi}\sigma}
    \exp\left(\frac{\tau^2}{\sigma^2}\right)
    K_0\left(\frac{\tau^2}{\sigma^2}\right)\simeq 0.23,\label{eq.hybrid}
\end{equation}
where $K_0(x)$ is the modified Bessel function of the second kind. The average is obtained by approximating $(v_L-v_R)$ as a Gaussian variable with standard deviation $\sigma$. The dependence on onsite disorder is shown in Fig. \ref{fig:peaks}, which shows a remarkable agreement between the numerical results and expressions \eqref{eq.IPR} and \eqref{eq.hybrid}. Converting these numbers back to frequencies in the mass chain gives: 
\begin{equation}
    \Delta f_0=\frac{\sqrt{k}}{4\pi\sqrt{m_1+m_2}}\Delta E\simeq 1.1 \text{ Hz}
\end{equation}
This value is close to the measured frequency difference of $f_6-f_5\simeq 0.9$ Hz.

The off-diagonal disorder was neglected, since off-diagonal disorder conserves the chiral symmetry of the topological edge state and does not contribute to the energy level separation, while onsite disorder breaks the chiral symmetry and shifts zero modes away from zero energy \cite{Scollon2020}. 

\section{Discussion and Conclusion}
The experiment demonstrates key features of the SSH model in a macroscopic, reconfigurable mechanical system. In relative-displacement coordinates, the system exhibits the off-diagonal dimerization characteristic of the SSH model. Most notably, the topological configuration supports boundary modes within the band gap that are localized predominantly at opposite ends of the chain. By contrast, the edge states of a disorder-free finite SSH chain form nearly degenerate symmetric and antisymmetric combinations, with amplitude at both boundaries.

The deviations from the ideal model are also physically informative. Mass variations shift the finite-chain eigenfrequencies, while the projection of non-longitudinal degrees of freedom onto the measured coordinate introduces additional perturbations. Because the ideal edge-state doublet is nearly degenerate, it is particularly sensitive to such perturbations. In a finite chain, weak onsite and hopping disorder can mix the symmetric and antisymmetric states, producing modes localized predominantly at individual edges without eliminating edge localization. Although the mid-gap modes remain in the topological configuration, disorder increases their frequency splitting and lifts their near-degeneracy \cite{Kvande2023}. These results show that SSH-like edge localization persists in the mechanical analogue despite weak onsite disorder. The system's robustness and reconfigurability make it a promising quantitative testbed for finite-size, disordered, and nonlinear phenomena in one-dimensional topological mechanics.

The authors acknowledge support from NSERC, INTRIQ and RQMP.

\bibliography{references}
\end{document}